PROPAGACIÓN DEL SARS-COV-2 EN CUBA. UNA VISIÓN CUALITATIVA DESDE LA TEORÍA DE SISTEMAS COMPLEJOS.

**Oscar Sotolongo-Costa,(1) Fernando Guzmán Martínez (2)**

1 Centro de Sistemas Complejos, Universidad de La Habana, Habana 10400. Cuba
2 INSTEC, Universidad de La Habana, Habana 10400. Cuba

**Abstract:** We describe some properties of the evolution of the COVID-19 pandemic that reveal its behavior as a complex system. The propagation mechanism shows a Poincaré section with fractal dimension $1 < D < 2$. The present period of apparent recovery shows a relaxation with non-extensive statistical properties.

**Resumen**: En este trabajo se describen algunas propiedades de la evolución de la pandemia de COVID-19 que ponen de manifiesto su comportamiento como sistema complejo. El mecanismo de propagación muestra una sección de Poincaré con dimensión fractal $1 < D < 2$. El presente periodo de aparente recuperación muestra una relajación con propiedades estadísticas no extensivas.

## I.-INTRODUCCIÓN

No hay dudas de que la infección de COVID-19, una vez convertida en pandemia, ha transformado las vidas de todos los habitantes del planeta, la mayoría de las veces para mal. No se trata solo de las repercusiones que de una u otra forma tiene en la salud física y mental de las personas; también está impactando de manera dramática en la economía mundial, lo cual a corto plazo ya trae muy serias repercusiones. En Cuba, por razones bien conocidas, es aún peor.

Casi todos los países están en distinta medida afectados. A nadie escapa que la tarea que se plantea en el orden científico es muy compleja. Es multidisciplinar, transdisciplinar y cualquier adjetivo que designe una profunda imbricación entre las distintas ramas de la ciencia en aras de un objetivo común. Esto se menciona en [1-4].

El objetivo de esta contribución es mostrar que la complejidad de esta pandemia en las condiciones de Cuba muestra rasgos de caos en el sentido de que la variación diaria de casos es susceptible de analizarse mediante mapas de retorno en secciones de Poincaré, mostrando dicho mapa propiedades fractales, sin que parezca que el mapa cubra la sección (comportamiento totalmente aleatorio) ni se pueda fijar una curva determinista en el mapa de primer retorno, mostrando así una dinámica que muestra un comportamiento de caos "débil".

Además, se trasluce del análisis de la data que, aún sin que se pueda haber demostrado en rigor la existencia de "ley de potencia" en algunas distribuciones, no es descartable la existencia de correlaciones espaciotemporales de largo alcance y la posible ocurrencia de una disminución relativamente lenta de los casos diarios, aunque se hayan tomado medidas estrictas de control en todo el país. Entonces especularemos sobre el rol de la vacunación.

Aquí se presenta un punto de vista que al parecer no se ha tratado extensamente en la literatura. Un enfoque que permite comprender la naturaleza de esta pandemia y que pudiera ser útil en el análisis de la evolución de los casos.

Como la terminología comúnmente usada en sistemas complejos difiere de la utilizada comúnmente en epidemiología, dedicamos el epígrafe II a una breve explicación de los conceptos y términos necesarios para introducirse en este campo. El epígrafe III ya comienza trabajando con dichos conceptos, como complejidad, ley de potencias, ley de escala, caos, etc. En el epígrafe IV se ve directamente la relación de la ley de escala con la pandemia y en el V se aplica al caso de Cuba. Los epígrafes VII y VIII aplican los conceptos mencionados a la evolución de la pandemia en Cuba y pretenden ilustrar cómo la situación ha ido evolucionando de una situación de caos a una donde existen rasgos de control de la pandemia. En las conclusiones se muestra que la presente situación de disminución acelerada de los casos diarios que tiene como factor que entendemos determinante a la vacunación no tiene que ser irreversible y es necesario que dadas las correlaciones que esta pandemia exhibe a nivel mundial, es preciso controlarla a ese mismo nivel y mantener ingentes recursos y protocolos para continuar con esta evolución favorable. Aunque esto parezca obvio, las razones están detalladas en este artículo soportadas por la teoría de los sistemas dinámicos y con la solidez y la valoración de universalidad que proporciona la solidez de las construcciones científicas.

## II.- CONCEPTOS BÁSICOS

a Complejidad

En la naturaleza y la sociedad hay múltiples ejemplos de complejidad, y todos tienen la propiedad de que componentes muy simples interrelacionados (Llamémosles "correlacionados") dan lugar a un comportamiento global coordinado del sistema del cual forman parte [5,6]. Sirva como ejemplo el sistema inmune, donde las componentes simples que lo conforman (anticuerpos, citocinas, etc.) originan un comportamiento complejo para combatir virus y bacterias.

Esta propiedad de coordinación se presenta de alguna manera también en la economía, la Internet, los bancos de peces, etc., de forma que las componentes del sistema presentan un comportamiento coordinado hacia un fin. Las coordinaciones podemos entenderlas en muchos casos como "interacciones" y siempre son de largo alcance. No tienen el mismo comportamiento un gas ideal de partículas que solo interactúan durante los choques que un gas de partículas cargadas cuyas interacciones son de largo alcance (un plasma) y dan lugar a comportamientos radicalmente diferentes.

Las correlaciones entre las partes del sistema complejo no tienen que manifestarse solamente a través de fuerzas. También se pueden establecer correlaciones mediante el intercambio de información como ocurre en las migraciones, infecciones, fluctuaciones financieras, etc.

En los sistemas inmunes de todos los organismos lo esencial es que el conjunto de entidades que los componen actúan de manera cooperada para combatir los ataques.

b Caos

Este concepto está asociado con el determinismo y la sensibilidad extrema a los pequeños cambios del desarrollo de un sistema. (ver [6])

En los sistemas mecánicos es de esperar que fuerzas pequeñas produzcan cambios pequeños, pero a veces puede ocurrir que una pequeñísima fluctuación provoque grandes cambios. Por ejemplo, una moneda en equilibrio sobre su borde puede caer ante u pequeño soplo, la caída de una piedra desde una montaña puede generar un gran alud, y así hay muchos ejemplos más.

De nuevo señalamos que los factores que provocan los cambios no tienen que ser fuerzas en el sentido ordinario de la palabra. Los factores pueden ser de otro tipo. Lo esencial es que una variación pequeña puede traducirse en cambios grandes de comportamiento. Sirvan de ejemplo algunos desórdenes mentales, crisis financieras y ya veremos que también la presente pandemia bajo determinadas situaciones.

Aunque para un sistema existan modelo matemáticos precisos y bien elaborados esto no significa que eso solo haga que el sistema se comporte siempre con perfecta regularidad, sea previsible y, por lo tanto, no sea caótico. Un ejemplo clásico es el clima, el cual no es pronosticable en detalle a largo plazo. Hablando en rigor, para conocer y predecir el comportamiento de un sistema caótico habría que conocer el estado inicial del mismo con infinita precisión, conociendo todas las variables que influyen en su comportamiento, de las cuales muchas no se consideran en el modelo matemático que se elabora por su despreciable influencia en el movimiento del sistema (por ejemplo, no necesitamos conocer la latitud del lugar en que estamos para saltar un charco, ni conocer la tensión superficial del agua para beber un vaso), sino además dar las condiciones iniciales de las variables relevantes con infinitas cifras significativas. Por eso la teoría del caos ha elaborado herramientas que nos permiten caracterizarlo, en los próximos epígrafes abundaremos en esto.

c Criticidad

Sirva un ejemplo de fenómeno crítico: Como sabemos los gases se pueden licuar si a una temperatura suficientemente baja se comprimen a alta presión. Pero a cierta temperatura, llamada “temperatura crítica”, al comprimir un gas hasta una cierta “presión crítica” deja de existir cualquier diferencia entre los estados líquido y gaseoso. En ese punto, llamado “punto crítico” el gas no se convierte en líquido, como ocurre a temperaturas menores que la crítica, sino en algo que ya no es gas pero no tiene tampoco las características de líquido.

Lo interesante de esto es que lo que se observa en la cámara de compresión ya no tiene la transparencia del líquido o gas, sino de una fase que sólo existe en el punto crítico y que muestra la propiedad de “opalescencia crítica”. El gas deja de ser transparente. Es opaco y dispersa toda la luz de todos los colores. Si nos movemos fuera del punto crítico, el gas recupera su transparencia ya esté en fase gaseosa o líquida (ver [7]).

El fenómeno ocurre porque en el punto crítico las moléculas del gas forman clústeres, grupos moleculares de todos los tamaños posibles, agrupándose de forma que las ondas luminosas de todos los colores se dispersan, lo que convierte al gas en opaco.

Esto es sólo un ejemplo de lo que se conoce como “fenómeno crítico”. La criticidad es la capacidad de formar sistemas con escalas de correlación, como los clústeres que mencionamos en el gas, de todos los tamaños. Las escalas de correlación divergen. Este es el rasgo sobresaliente, y muy sobresaliente, de la criticidad. Donde hay criticidad la magnitud que se hace crítica, llamémosle $"y"$ depende de un parámetro $"x"$ o $"\text{x-x}_0"$ en la forma $y = x^{\pm a}$ o $|x - x_0|^{\pm a}$ o sea ley de potencias aquí “a” es el exponente crítico. (para un tratamiento muy completo de este comportamiento ver [7]).

Otros fenómenos críticos son la pérdida de la magnetización en los ferromagnéticos cuando la temperatura llega a la llamada “temperatura de Curie”, también el gigantesco aumento de la susceptibilidad magnética en la vecindad de esa temperatura y muchos otros fenómenos.

Una característica común de los fenómenos críticos es que al variar el parámetro que lleva los sistemas al punto crítico, la magnitud que se hace crítica (escala de los clústeres, susceptibilidad, etc.) depende del parámetro como una ley de potencia (también llamada muchas veces ley de escala).

Pero todavía más importante para nuestros propósitos es que en la naturaleza se dan “criticidades” sin que haya que sintonizar ningún parámetro. Son criticidades que se forman solas Esto se conoce como “criticidad autoorganizada”. Este concepto fue introducido por un trabajo pionero de P. Bak, C. Tang y K. Wiesenfeld en 1987 [8].

El ejemplo típico es la pila de arena. Podemos hacer una pila de arena dejando caer granos sucesivamente. Al principio, simplemente se va formando la pila por sucesiva aglomeración de granos hasta que la hacerse muy grande localmente la pendiente de la pila, un grano cae. Este puede desliarse hasta la base sin afectar a los demás, pero también puede arrastrar otros en su caída formando una avalancha. La pregunta es: ¿De qué tamaño es la avalancha? O mejor, ¿De qué tamaño es la avalancha más probable?

La respuesta es: No existe una “avalancha más probable” La distribución de tamaños de las avalanchas es una ley de potencias, o sea la gráfica no muestra un pico en ningún valor siendo que las pequeñas avalanchas son más probables que las grandes.

Ya sabemos entonces que las avalanchas son fenómenos críticos autoorganizados.

## III.- LEY DE ZIPF

La ley de Zipf, debida al lingüista George Kingsley Zipf, se originó en el estudio de la distribución de las palabras en el idioma inglés. Según esta ley, la frecuencia de ocurrencia de palabras en cualquier obra escrita es inversamente proporcional al “rango” de dicha palabra en frecuencia. Se define “rango” una magnitud ordinal proporcional al inverso de la abundancia. Es decir, la palabra más abundante tiene rango 1, la siguiente rango 2, y así sucesivamente. Esto vale también para los demás idiomas [9,10].

En efecto, dada la definición de rango, se hace obligatorio que las palabras menos frecuentes tengan mayor rango y a la inversa, pero lo interesante es que en una gráfica logarítmica los puntos de la distribución se agrupen a lo largo de una línea recta [10]. Eso es, precisamente, la ley de escala. Si en una ley matemática una magnitud $"y"$ depende de la variable independiente como una ley de potencia del tipo $y = x^n$, al menos en cierto rango de valores, diremos, aunque no sea del todo riguroso, que existe una “ley de escala”. Aquí $"x"$ es la variable independiente y $"n"$ un número llamado “exponente crítico”, como ya se dijo en el epígrafe anterior.

Así, la ley de Zipf es una ley de potencia entre el rango y la abundancia. O sea, si llamamos $"r"$ al rango y $"N"$ a la abundancia, se cumple que

$$N = r^{-a} \qquad (1)$$

En este caso el signo negativo aparece porque en el caso de la ley de Zipf a mayor rango menor abundancia. Expresada en términos de logaritmos, la ecuación (1) adopta la forma:

$$\log N = -x \log r \qquad (2)$$

A diferencia de (1) la gráfica de (2) es una recta con pendiente negativa entre las variables $\log N$ y $\log r$.Las ecuaciones (1) y (2) expresan la misma ley, como se ve, y en la literatura especializada ambas son empleadas a conveniencia.

En economía, la ley de Zipf se traduce en la “ley de Pareto”, que revela una relación de potencia entre el número de personas que reciben una cierta entrada monetaria y el monto de esa entrada. Ya aquí no se introduce el rango, La ley de Pareto se expresa como una distribución de frecuencia [10]. Curiosamente, de forma algo sobre simplificada se comenta que la ley de Pareto revela una relación 80/20, o sea, el 80% de la población solo recibe el 20% de la riqueza.

La ley de Zipf se aplicó rápidamente a la distribución de población en las ciudades: El rango de una ciudad es inversamente proporcional al tamaño de su población. La ciudad más poblada de un conjunto tiene rango 1, la segunda en orden rango 2, etc. Esto es posiblemente conocido por el lector, pero sorprende que esta ley sea también aplicable a la distribución de tamaños de meteoros, ingresos, campos de petróleo, etc. Lo que se ve entonces es que en estos casos existe una ley de potencias vinculando una magnitud con otra.

Aquí veremos la aparición de esta ley en algunas variables aplicadas a la pandemia.

## IV.- EL COVID Y LAS LEYES DE ESCALA.

Las leyes de tipo Zipf expresan una propiedad general de los sistemas que es independiente de su composición y de los detalles del sistema.

Por ejemplo, todos estamos acostumbrados a que una serie de mediciones de cualquier magnitud física, pongamos por caso la longitud de una mesa ordinaria, se distribuyen siguiendo una ley Gaussiana (Normal) donde los datos se distribuyen en forma de campana con un máximo en el valor correspondiente al valor medio de la serie de mediciones.

Pero ya tenemos incorporado que así debe ser; lo que pasa en la longitud de una mesa vale para la estatura de una persona, el peso, el volumen de sangre bombeada por las personas a tal edad, su temperatura, etc., solo por poner ejemplos médicos. Es una creencia general que todas las magnitudes a nuestro alrededor se distribuyen en torno a un valor medio. La distribución estadística de valores debe distribuirse cerca de un valor más probable, de modo que los valores muy lejanos tanto por encima como por debajo del mismo deben ser muy escasos y aparecer con muy poca frecuencia. En una ley de escala como las que hemos mencionado esto no ocurre, sino que los valores más pequeños son siempre más abundantes que los mayores. NO hay un valor más abundante que los demás al menos en los rangos que se consideran en el fenómeno. Desde la época de Da Vinci se asociaron valores característicos, algunas veces identificados como valores óptimos, sea por razones estéticas o de salud, a esas magnitudes. Quizás la famosa proporción 90-60-90 tan mencionada en los certámenes de Miss Universo, venga de esa corriente.

Pero las cosas no son tan simples y ya vimos el ejemplo de la ley de Pareto-Zipf, y en general de distribuciones con leyes de escala. Y no se limitan a uno o varios campos, tienen una gran ubicuidad. Ejemplos:

-Fluctuaciones de la bolsa.

-Distribución de la energía de los terremotos.

-Distribución de los cráteres lunares

-Distribución de los meteoritos por tamaño

-Distribución de los diámetros de los capilares, de los conductores de xilema en los árboles, de los tamaños de las islas, etc.

Ahora queda saber si en la pandemia de COVID-19 hay comportamientos de tipo ley de escala.

**V.- SOBRE LA PANDEMIA EN CUBA.**

La pandemia de COVID-19 se manifestó en Cuba a partir del 11/3/2020 reportándose diariamente el número de contagios, hospitalizados y fallecidos. La evolución de esta pandemia ha sido como sigue:

La fluctuación relativa en el número de casos diarios, definida como el valor absoluto de la relación entre la variación de casos entre un día y el siguiente dividida por el número de casos de este último día se distribuye como se muestra en la figura 1.

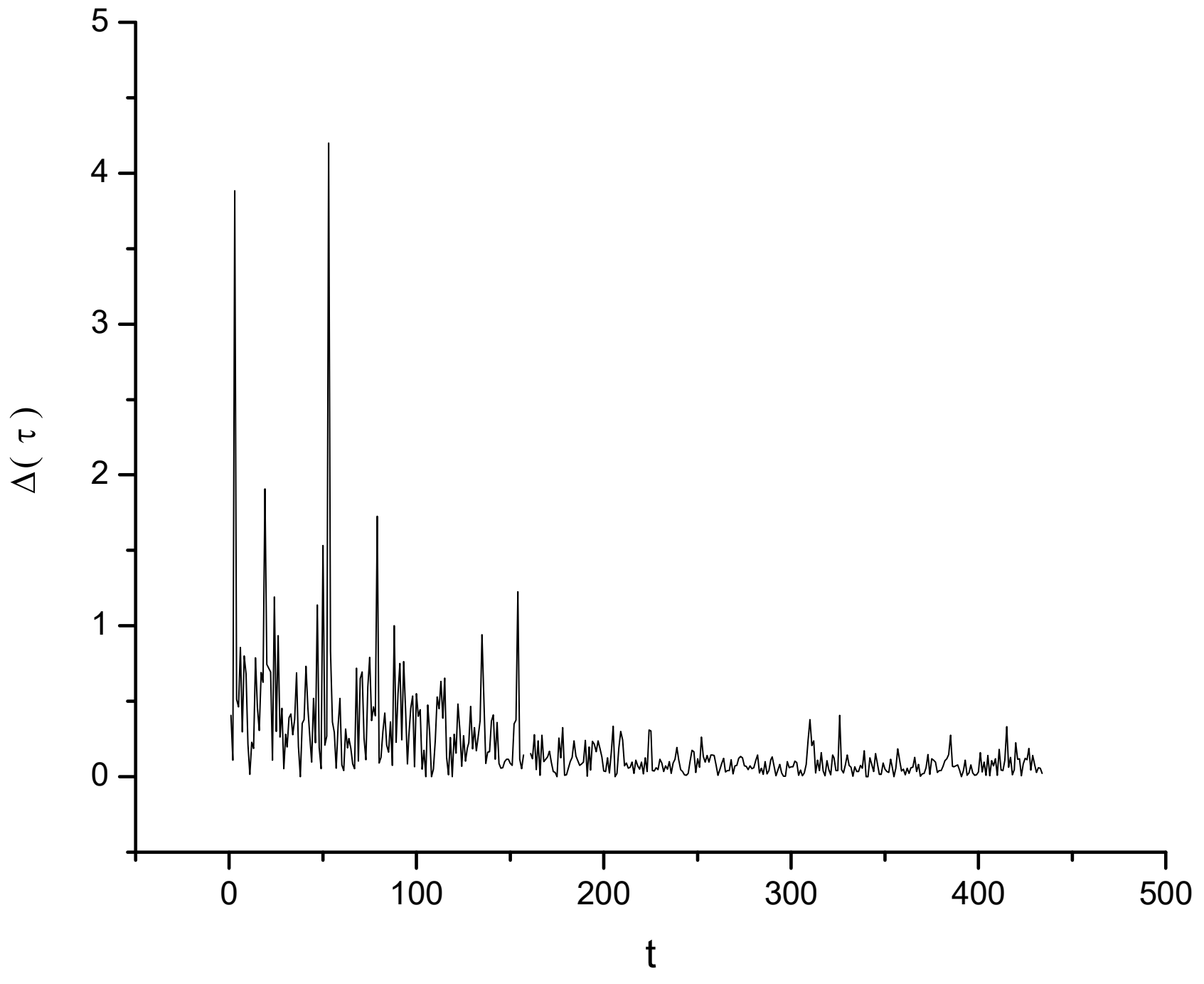

Fig.1: Variación temporal de la fluctuación relativa desde el 17/8/21 al 24/10/21

Cabe preguntarse cómo se distribuyen las fluctuaciones relativas, es decir, qué probabilidad hay de ocurrir una fluctuación relativa alta o baja. Los datos nos dan lo siguiente:

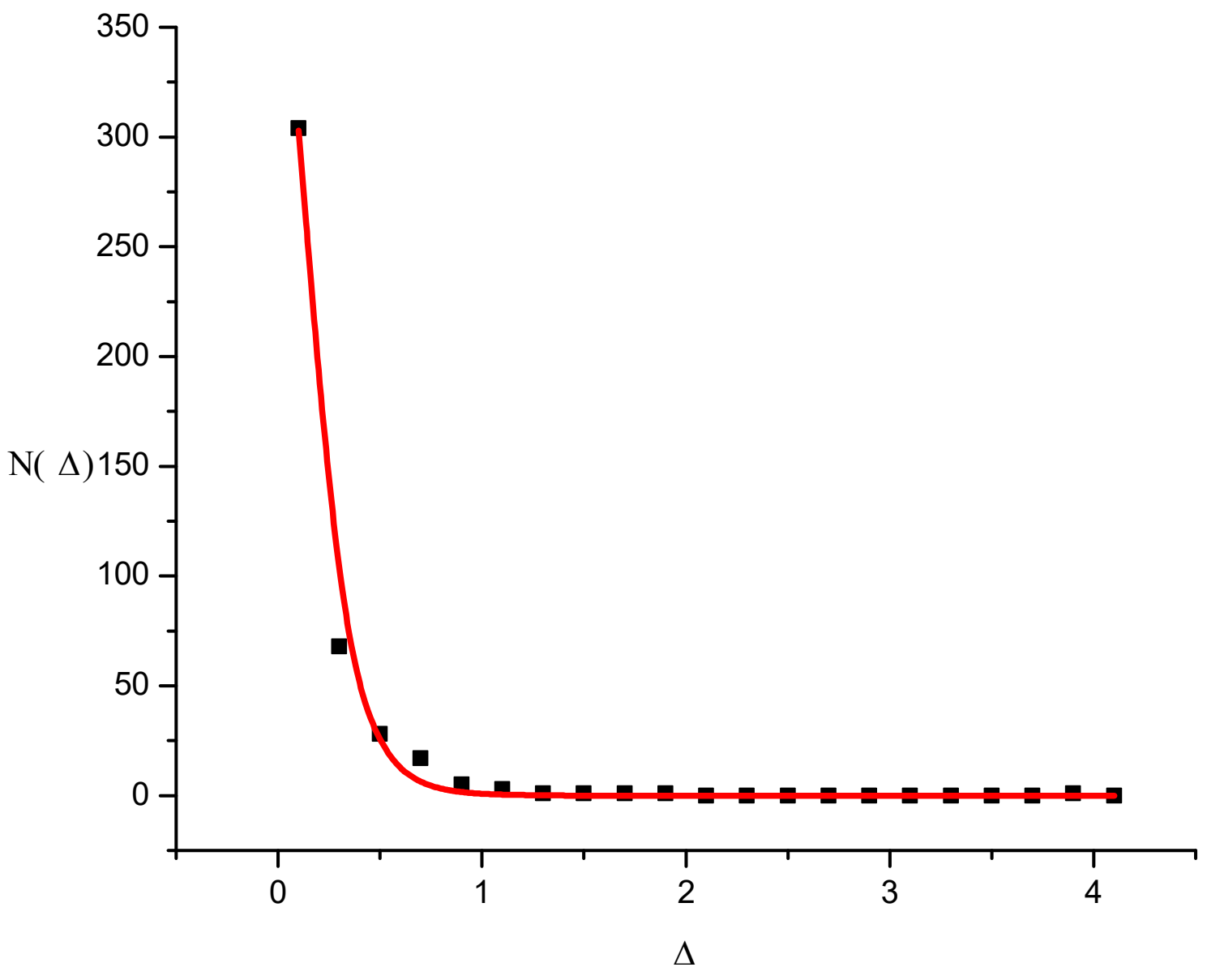

Fig.2: Distribución de las fluctuaciones relativas hasta el 24/8/21. Responde a una distribución exponencial decreciente $N(\Delta) \sim e^{-\Delta}$, es decir, la probabilidad de fluctuaciones grandes es pequeña, y a la inversa. Esto es lógico ya que con el tiempo los casos han ido aumentando y las variaciones están referidas al número de casos. Sin embargo, no es una distribución que manifieste ley de escala ya que la función de distribución es una exponencial, no una ley de potencia.

Pero por ahora no hemos señalado las leyes de escala ni otras propiedades también importantes, en particular referentes a lo que pudiéramos llamar la "geometría" de la pandemia.

## VI.-INVARIANCIA DE ESCALA Y LEYES DE ESCALA

La existencia de distribuciones con ley de escala caracteriza a procesos omnipresentes en los comportamientos complejos, sistemas caóticos y con existencia de correlaciones espaciotemporales de largo alcance. Es obvio que, siendo la COVID-19 un fenómeno global las correlaciones entre las partes del sistema deben ser de largo alcance. En efecto, los contagios no precisan necesariamente de un solo factor, como pudiera ser un "vector" de transmisión, ni exige mucha cercanía entre el portador del virus y el receptor. Puede también ocurrir que un portador residente a gran distancia de otras personas, dada la movilidad, pueda en un viaje transmitir el virus en un lugar lejano. Los aerosoles, los objetos, etc. pueden también ser agentes de transmisión.

Por eso, veremos la dinámica de la pandemia a través del método conocido como "mapa de Poincaré" graficando el número de casos en Cuba el día n+1 en función de ese número el día n. Tenemos un mapa de primer retorno que contiene los pares ($X_n$, $X_{n+1}$ ) desde el 17/8/2020 hasta el 26/10/21.

Hagamos el gráfico de primer retorno para los primeros doscientos casos a partir del 17/8/20. Y veamos qué figura se obtiene.

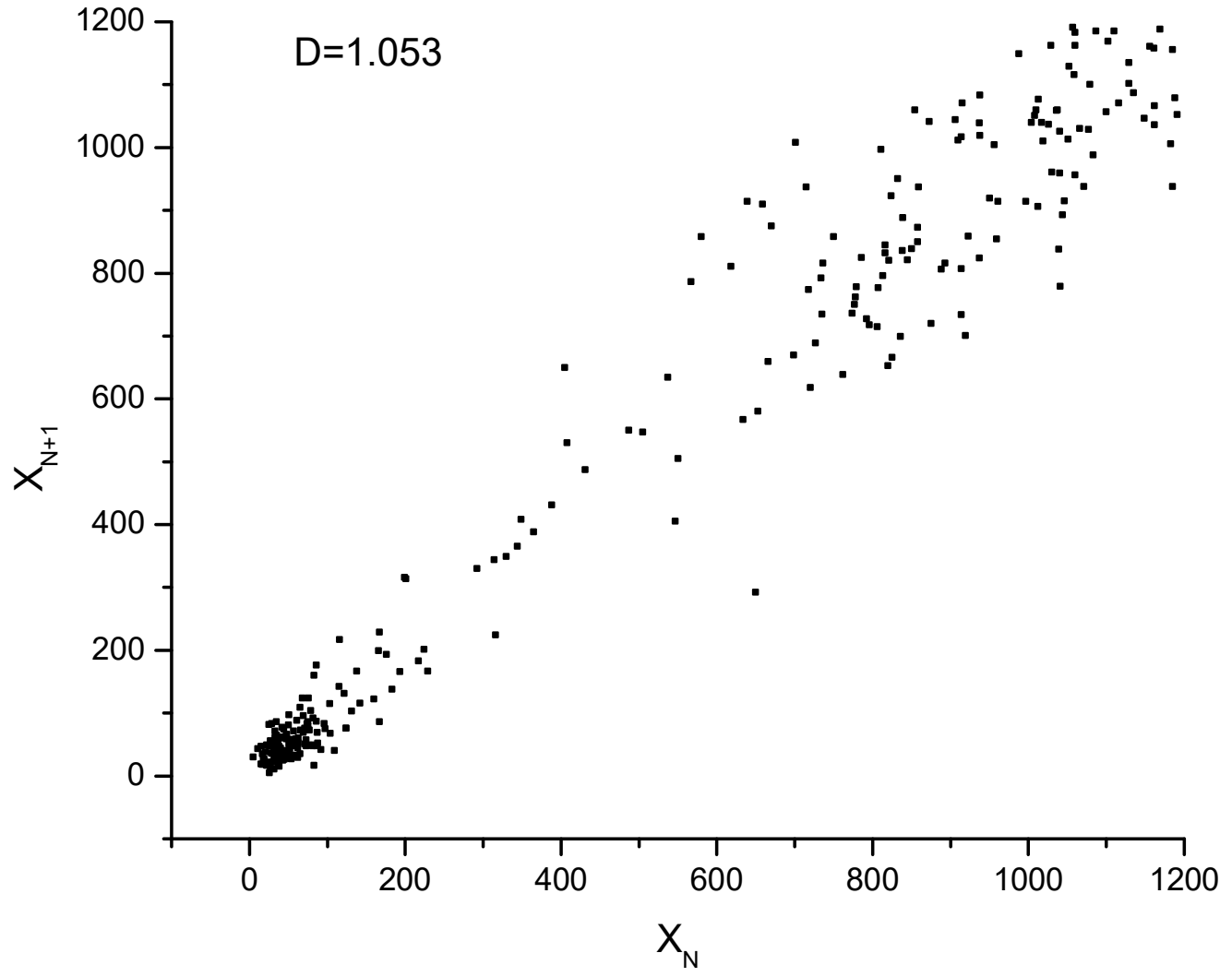

Figura 3.- PRIMEROS 200 CASOS

Aquí se ve una colección de puntos dispersos con una forma determinada. En el eje de las accisas se sitúa el número de casos un día mientras en el de las ordenadas el número correspondiente el día siguiente, así sucesivamente dando una figura compuesta por los pares ($X_n$ , $X_{n+1}$ ). Veremos que esta figura se caracteriza por su autosemejanza, o sea que en la escala de los 1000 casos, 2000 y hasta 10000 casos la figura no cambia sensiblemente su forma. Es, como se llama, invariante de escala. Las figuras con esa propiedad se conocen como “fractales” y los caracteriza el tener una dimensión no entera, llamada dimensión fractal, a diferencia de las figuras ordinarias que pueden ser unidimensionales o bidimensionales. Los detalles sobre las propiedades fractales de diversos sistemas pueden verse en [11]. El mapa de primer retorno muestra una figura con una dimensión fractal de 1.053, que es mayor que 1 y menor que 2, por lo que no caracteriza ni a una curva, que es una figura unidimensional, ni a una superficie, que es bidimensional. Cae dentro de los conocidos “fractales”, sistemas que muestran de alguna manera una dimensión fraccionaria y no pueden describirse por la geometría ordinaria euclidiana. Pero veamos qué figura corresponde al llevar los casos a, un instante posterior, digamos, 312.

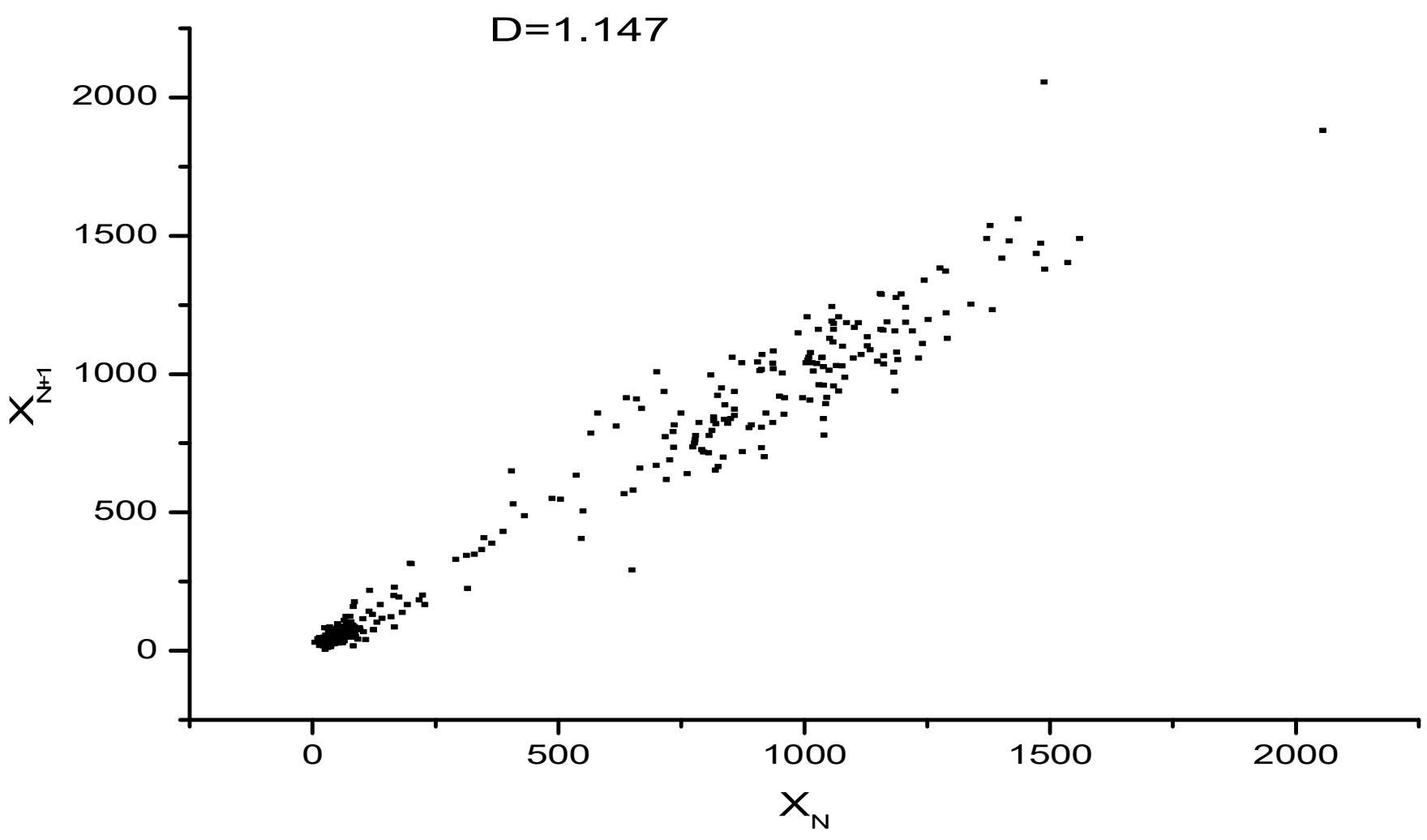

Figura 4.- PRIMEROS 312 CASOS

Resulta una figura que se asemeja mucho a la primera, aunque no es idéntica. La dimensión fractal es cercana.

Y si llegamos a los 340 casos:

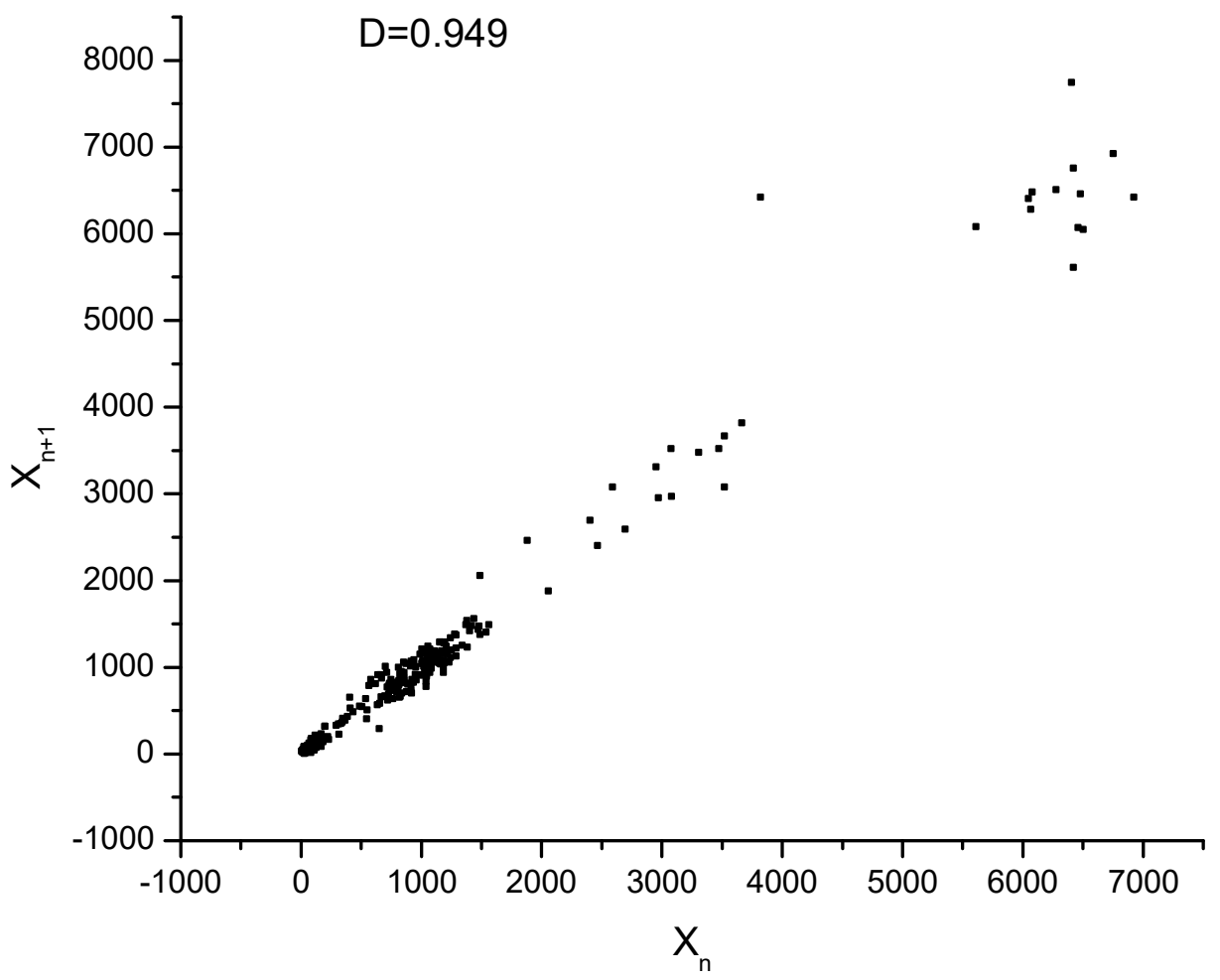

Figura 5.- PRIMEROS 340 CASOS.

La figura se "afina" lo que significa un aumento considerable en el número de casos. Ya el sistema se moverá entre los 6000 y 10000 casos.

Ahora veamos la figura correspondiente al 26 de octubre de 2021.

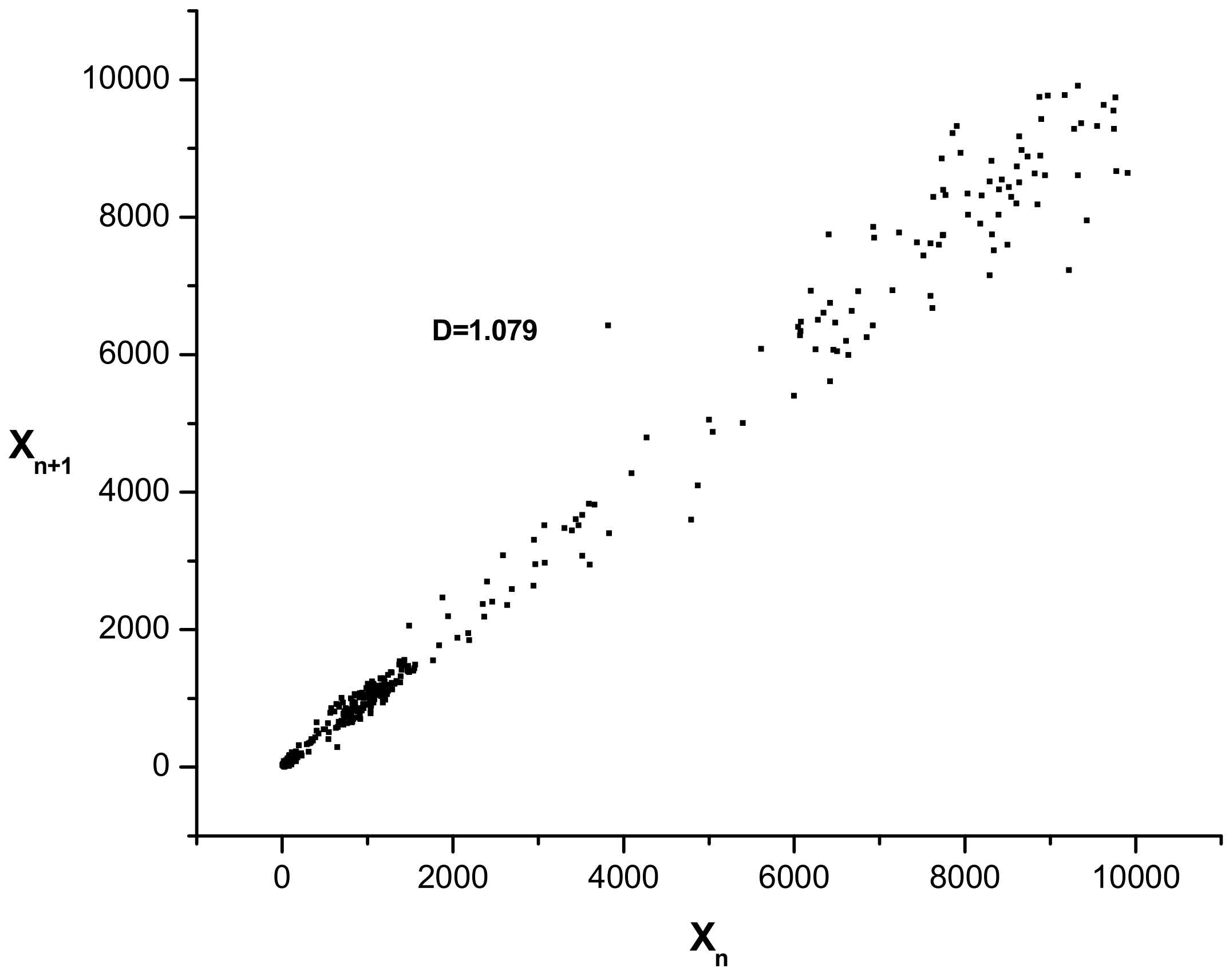

Figura 6.- OCTUBRE 26/2021

La figura 3 tiene una escala correspondiente al orden de 1000 casos, la segunda ya abarca una escala del orden de los 2000, o sea, el doble y ya la de la figura 3 del orden de los 10000. A pesar de los evidentes cambios de escala, puede apreciarse semejanza de la forma de todas ellas.

Un físico diría que se muestra una invariancia de escala, al menos aproximada. Así ocurre en la naturaleza. El diseño de los ríos, las nervaduras de las hojas, las ramas de los árboles, el sistema circulatorio, etc., etc.… Además, la dimensión fractal no muestra grandes cambios.

La figura es una distribución de puntos con dimensión fractal variable en el tiempo que tiende a un valor límite de $D \simeq 1.13$.

Si observamos el extremo inferior vemos una aglomeración de puntos muy grande a diferencia del extremo superior en que los puntos están muy separados. Esto es consecuencia de la dinámica del sistema, en que se determina una distribución en que la distancia entre un punto y el siguiente viene dada por la distribución de fluctuaciones relativas que están determinadas por el valor que representan los puntos contiguos en el tiempo. Esa propiedad se conserva en cualquier región del plano de Poincaré, o sea a cualquier escala. Es decir, la figura es invariante de escala. De ahí la naturaleza fractal del mapa. Hay invariancia de escala en las distintas regiones.

En la zona n<200 puede hacerse una valoración aproximada de los mapas de primer retorno. Para distintos intervalos de tiempo vemos que los puntos no tienden a definir una curva, sino a cubrir una región del plano.

Así podemos considerar esa región como de dimensión mayor que 1 y menor que 2. O sea, el mapa completo no cubrirá el plano.

Lo hecho en la figura 6 es equivalente a llevar las secciones de Poincaré correspondientes a distintas fechas a un tamaño similar. Se ve que lo que caracteriza esas distribuciones de puntos es la dimensión fractal, que como se ve resulta similar en secciones de Poincaré de distinto tamaño. Precisamente la dimensión fractal nos sirve en este caso para comparar unas distribuciones "desordenadas" de puntos que a pesar de tener formas algo parecidas nos sería difícil decir qué se conserva aproximadamente igual en todas las escalas.

En efecto, al medir la dimensión fractal de este mapa usando el programa ***Fractalyse*** para todos y cada uno de los días, se halla que la dimensión fractal varía poco excepto cuando había pocos puntos, o sea en los primeros días de la pandemia. Pero a esta altura la variación de D ocurre en las proximidades de la unidad y el proceso de ajuste muestra que tiende asintóticamente a $D \approx 1.13287$. Esto dice que el sistema se mantendría en una región finita del plano de Poincaré dando lugar asintóticamente a un fractal de esa dimensión.

Es posible continuar presentando datos que demuestran que en la dinámica de la epidemia hay correlaciones de largo alcance y presencia de geometría fractal; además puede mostrarse que hay alometría entre las distintas regiones del país, lo que demuestra una interdependencia que trasciende cualquier medida de aislamiento tomada.

Todo lo anterior argumenta a favor de la *existencia de complejidad y caos en la pandemia*. Si la dinámica de la pandemia estuviera regida por una ecuación explícita función del tiempo, el mapa de Poincaré sería una curva unidimensional. Por el contrario, si fuera un proceso completamente aleatorio los puntos en el mapa se distribuirían cubriendo toda el área accesible del plano tendiendo a formar una superficie bidimensional. La característica principal del caos radica en su sensibilidad extrema al cambio de condiciones iniciales. Traducido a este caso significa que un pequeño cambio o alteración del cumplimiento de los protocolos de aislamiento y seguridad, o de los insumos adecuados para el cumplimiento de los protocolos, etc., puede influir muy fuertemente en el desarrollo de los contagios.

Si a esto agregamos que el virus ha generado una variedad de cepas con distintas características de contagiosidad, mortalidad, y resistencia a las vacunas, y que además estamos en una etapa en que debido al surgimiento de una nueva cepa (Delta) han aumentado mucho los fallecimientos, distribuyéndose entre todos los rangos de edades y tipos de comorbilidades, incluyendo aparentemente fallecidos sin comorbilidades, la aserción de que pequeños cambios, -en este caso una mutación- pueden inducir grandes consecuencias, adquiere una fuerza aún mayor.

La invariancia de escala en la geometría de la sección de Poincaré nos ha servido para caracterizar el desarrollo caótico de la pandemia en Cuba. Podemos también referirnos a la invariancia de escala explícita en leyes de potencia de otras magnitudes. Esto no es nuevo pues ya en [12] se señala la existencia de leyes de escala en la propagación del COVID-19. No obstante, aquí mostramos otras relaciones.

## VII.-LAS CORRELACIONES Y SUS CONSECUENCIAS.

La pandemia es un proceso de interacción con correlaciones temporales y espaciales de largo alcance. La población es el sistema que posee estas correlaciones, y se puede usar la entropía de Tsallis en este caso para describir la variación diaria de estos.

La expresión para la entropía de Tsallis es [13]:

$$S_q = \frac{1-\sum_{i=1}^{W} p_i^q}{q-1} \tag{3}$$

En unidades de la constante de Boltzmann $k_B$. La magnitud $p_i$ es la probabilidad de que el sistema en estudio se encuentre en el estado $i$. El parámetro $q$ se conoce como "parámetro de no extensividad" y puede demostrarse que cuando $q \to 1$ esta entropía se reduce a la conocida entropía de Boltzmann-Shannon $S = \sum_i p_i \log p_i$.

Usando (3), Normalizando las probabilidades y haciendo el cálculo de la probabilidad de estado para la magnitud continua x, tenemos en el caso continuo una expresión del tipo.

$$p(x) = A\left[1+B(q-1)x^2\right]^{\frac{1}{1-q}} \tag{4}$$

A y B son constantes a ajustar con los datos experimentales. Los detalles de esta entropía pueden verse en [13] y sus referencias.

La forma matemática es de (4) es una forma generalizada de la exponencial dependiente del cuadrado de la variable, en definitiva, una generalización de la ley estadística de Gauss que puede usarse para ajustar las curvas de casos diarios. Se ha bautizado como “q-Gaussiana”.

Ahora, volviendo a la relajación del sistema, podemos tratar de predecir cómo será. Para ello, acudiendo a la teoría de la relajación aplicada en [14] podemos dividir el sistema en clústeres de distinta magnitud, Cada uno caracterizado por un tiempo de relajación inverso propio $\beta$, que relaja de manera clásica según $e^{-\beta t}$ pero la función de relajación del sistema será la ponderación de la de todos los clústeres. Así:

$$\Phi(t) \sim \int_0^{\infty} f(\beta) e^{-\beta t} d\beta \qquad (5)$$

Si suponemos que el tiempo de relajación inverso es función potencial de la magnitud del clúster $\beta(x) \sim x^{\alpha}$, y que la densidad de probabilidad $f(\beta) \sim x^{\beta}$, donde $x$ es la talla del clúster, entonces (5) da:

$$\Phi(t) \sim \frac{t^{\frac{1+\alpha}{\beta}} \Gamma\left(-\frac{1+\alpha}{\beta}\right)}{\beta} \sim t^{\gamma} \qquad (6)$$

Siendo $\Gamma(x)$ la función gamma de Euler.

O sea, la relajación es según una ley de potencia del tiempo si el sistema es no extensivo. Una relajación lenta, con ley de escala. No obstante, el *comportamiento real de la pandemia no muestra una relajación lenta*.

Pero como ya se dijo, esto ocurriría si los clústeres se distribuyen según una ley de escala, una ley de potencia del tamaño.

Podemos comprobar si para la pandemia esto puede darse. Dividamos el país en clústeres que se identifiquen con las provincias, siendo sus tamaños la cantidad de casos reportados en un día. Claro que con esto cambian los “rangos” que ocupen las provincias de día a día, pero no importa si los rangos de las provincias en cada momento se acomodan a una recta en la escala doble logarítmica. En efecto, la figura 10 demuestra que en una etapa temprana hay ley de Zipf entre las distintas provincias.

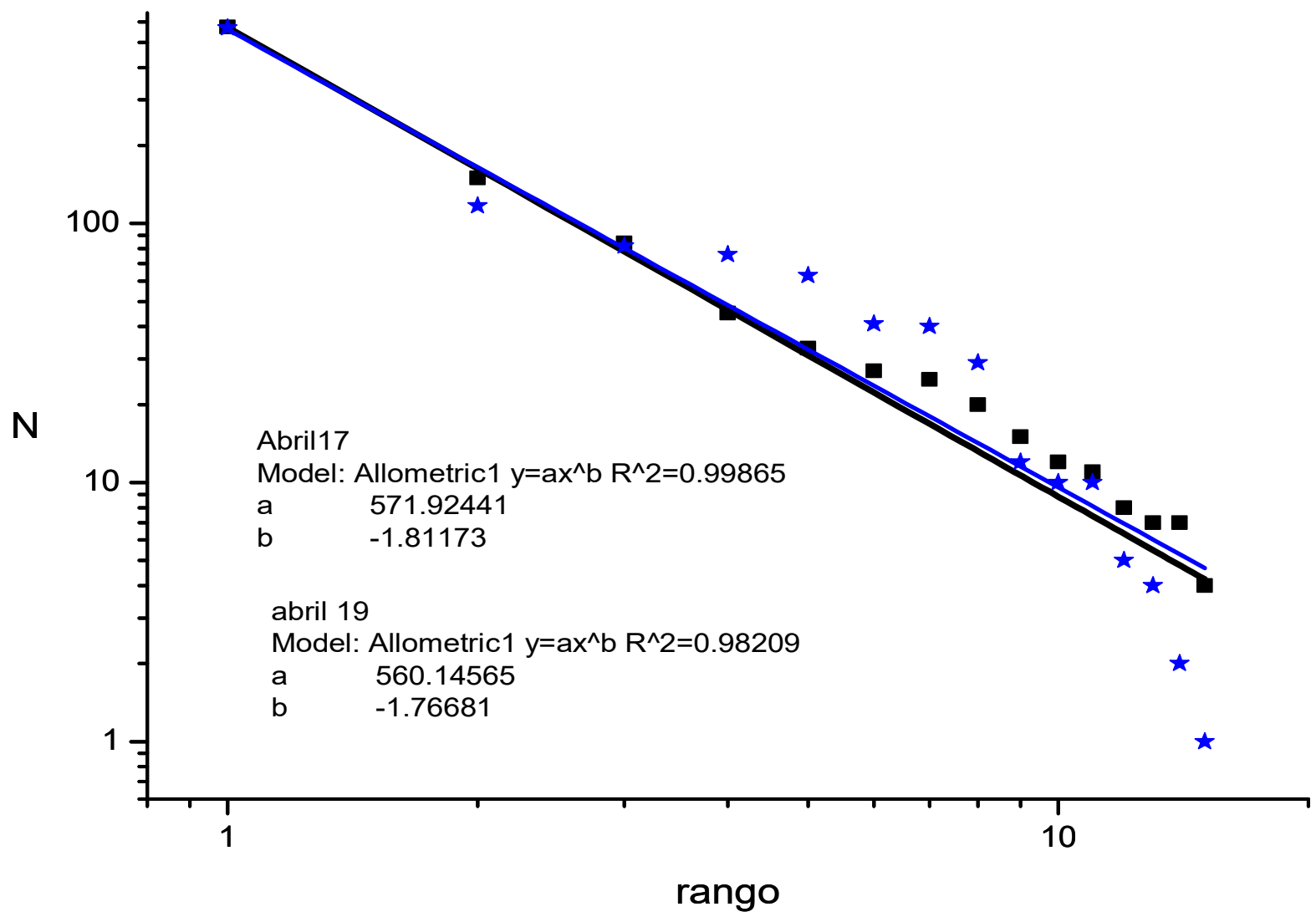

Figura 7.- Disposición de las provincias en una gráfica log-log del número de casos del día 17 de abril. (en negro) con coeficiente de correlación $R^2 \simeq 0.99$ y abril 19 (azul) con $R^2 \simeq 0.98$. La ley de Zipf se cumple muy bien La pendiente en ambos casos revela que la ley de distribución de clústeres en este caso es una ley de potencia del tipo $f[x] \sim x^{-2}$.

Para ejemplificar mejor cómo se ve la ley de potencias, podemos tomar la misma data, por ejemplo, de Abril17 y hacer la gráfica en escala lineal. La data se dispone en una hipérbola como la de la figura 8:

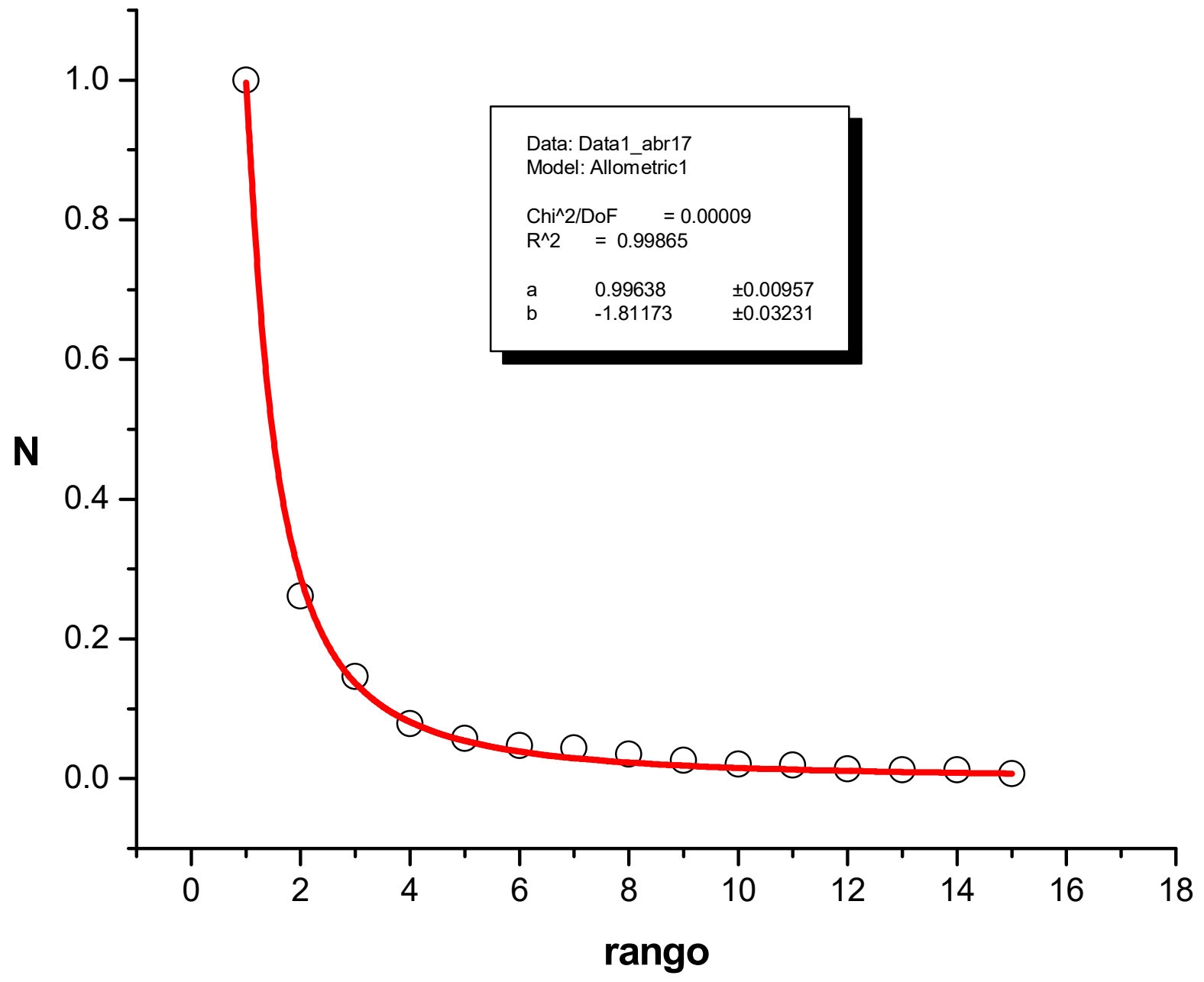

Fig.8.- Gráfica de la alometría entre los casos de COVID-19 en abril 17 2021. Esta gráfica representa lo mismo que la figura 7 y como se ve, no cambian los parámetros de la distribución. El exponente crítico sigue siendo -2. Lo única que varía es que los ejes coordenados están en escala lineal y los de la figura 7 en escala logarítmica.

Es de esperar que, si la relajación ocurriera en estas condiciones, o sea con las medidas que hay hoy, con una ley de Zipf para las provincias, pero sin la vacuna, sería un proceso lento. El hecho de que la relajación en estos momentos no sea lenta obedece en nuestra opinión a la acción de un proceso externo modificador de la dinámica del sistema, el proceso de vacunación. La aplicación de las vacunas a gran escala evidentemente modifica el comportamiento de la propagación. Si bien durante el proceso desde inicios del 2020 hasta julio/2021 cuando aún la influencia de la vacunación era débil en el país, la pandemia fue autoorganizándose hacia un estado crítico donde las leyes de potencia y la invariancia de escala permitirían hablar de “criticidad autoorganizada” de la COVID en Cuba; la influencia de la vacunación masiva introdujo una ruptura en la dinámica, de modo que la relajación muestra una caída más bien rápida de los casos diarios y parece depender del tiempo ajustándose bien con una relajación rápida. Lo que nos lleva a preguntarnos si la dinámica caótica, las correlaciones de largo alcance y las leyes de escala han sufrido transformación en fecha reciente.

Busquemos la respuesta en la propia relajación, que ya sabemos que no es lenta. Si el sistema ha superado la etapa de scaling puede probarse si el modelo de relajación propuesto en [14], que resultó bueno al inicio de la pandemia para describir la propagación en varios países, puede ajustarse aquí. La ecuación obtenida en [15] basada en un modelo de derivada fractal tiene una forma muy parecida a la q-gaussiana. En efecto, El número de casos diarios $n(t)$ en ese modelo obedece a la ecuación:

$$n(t) = A\left[1+(1-q)B(t^{\alpha}-t_0^{\alpha})\right]^{\frac{1}{1-q}} , \qquad (7)$$

donde $A, B, \alpha$ son el número inicial de infecciones $n_0$, la combinación $\frac{1}{\tau^{\alpha} n_0^{1-q}}$, siendo $\tau$ un tiempo de relajación característico introducido con objeto del ajuste y el exponente "fractal" u orden fractal de la derivada del número de casos en el tiempo. $q$ es el parámetro de no extensividad de Tsallis, como ya se explicó. Aquí se tomó el instante inicial (o sea el instante en que comienza la relajación, como cero, y para observar la concordancia con la q-gaussiana se tomó $\alpha = 2$.La figura 12 muestra la concordancia para la relajación con el modelo fractal propuesto en [15].

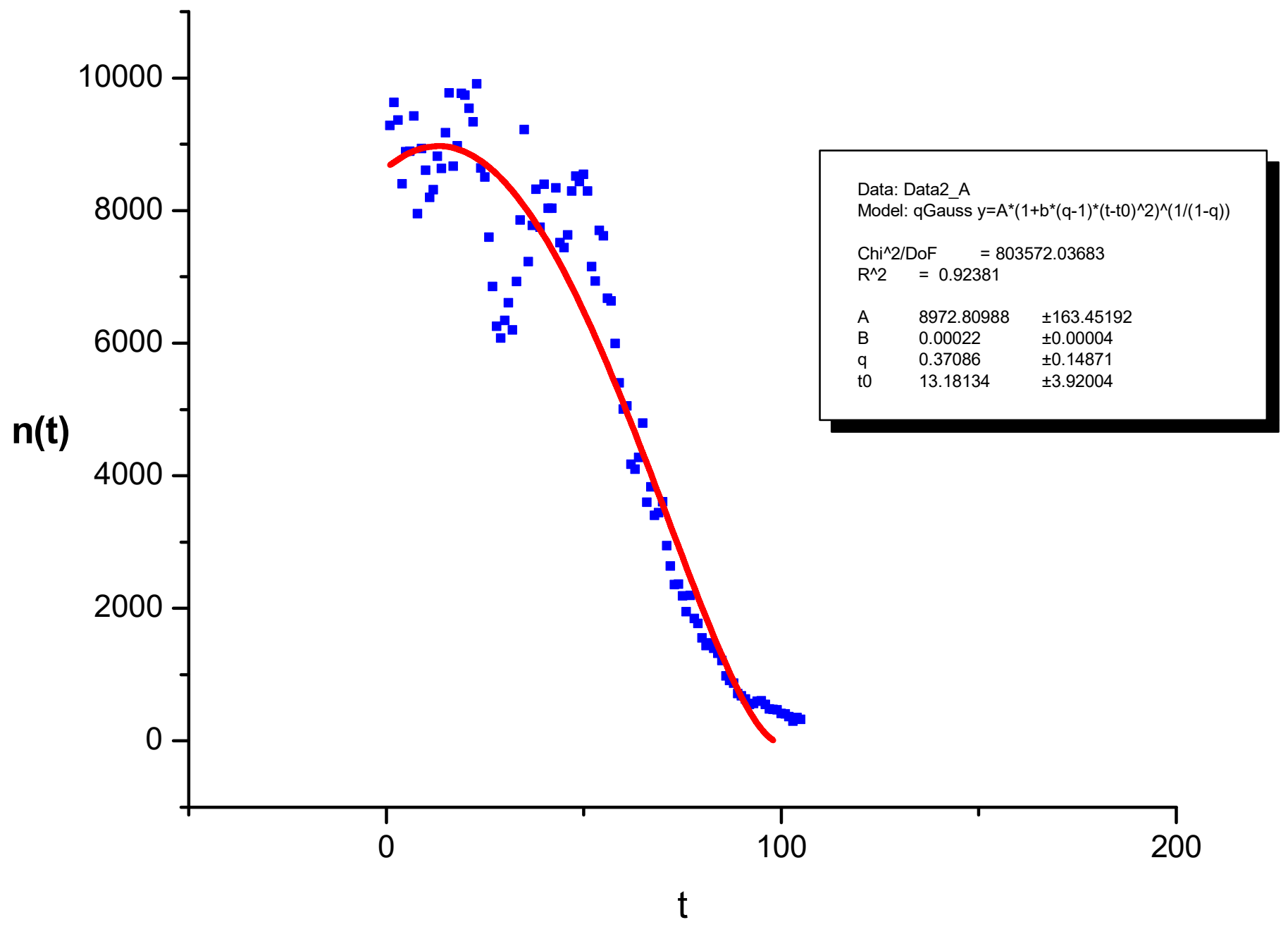

Fig.9.- Ajuste de la data de la cola de relajación del número de casos diarios con una q-gaussiana. La data cubre del 2/8/21 al 15/11/21.

Como se ve, el ajuste es bueno. También hemos comprobado que una relajación exponencial decreciente no ajusta tan bien como este modelo con la data y que una cola larga de tiempo no da buen ajuste. En estos momentos el sistema se comporta de modo

“menos complejo”. Se ha debilitado la correlación de largo alcance. Tienen que haber dejado de valer las condiciones deducidas en [14] para la relajación lenta.

Comprobemos la alometría entre las distintas provincias. En efecto, la ley de Zipf ya no ajusta tan bien. (fig.10)

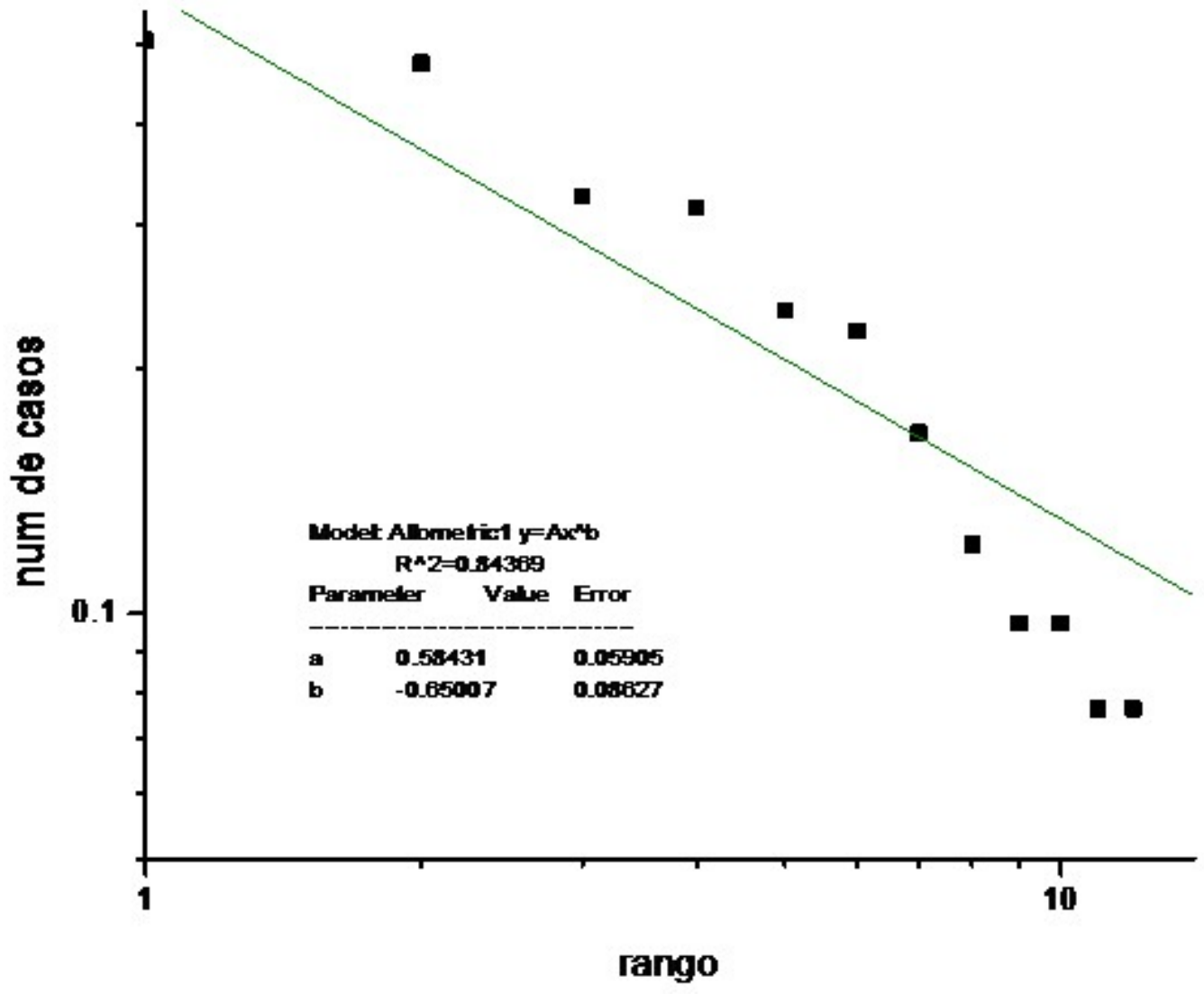

Fig.10.- Disposición de las provincias por rango de contagio. Datos del 15/11/2021. Observar el debilitamiento del ajuste con la ley de Zipf. Aquí igual que en la figura 10, la representación puede hacerse en escala lineal sin que cambie nada.

Ahora las correlaciones son más débiles que en fechas anteriores. La presencia de un factor externo perturba al sistema impidiendo una autoorganización hacia el estado crítico en que se iba formando. Antes la entropía (de Tsallis) aumentaba más fuertemente al no haber un factor externo que rompiera su aislamiento y sus correlaciones predominaban. La vacunación es, al parecer, el factor que impide el paso a la criticidad autoorganizada al cambiar la naturaleza de los componentes del sistema

Además, si la dinámica del sistema ha cambiado, debe observarse en la dimensión fractal del plano de Poincaré. Analicemos la misma para los casos desde el 2/8/21 al 15/11/21, los 104 días del 2/8/21 al 15/11/21. La misma se presenta así:

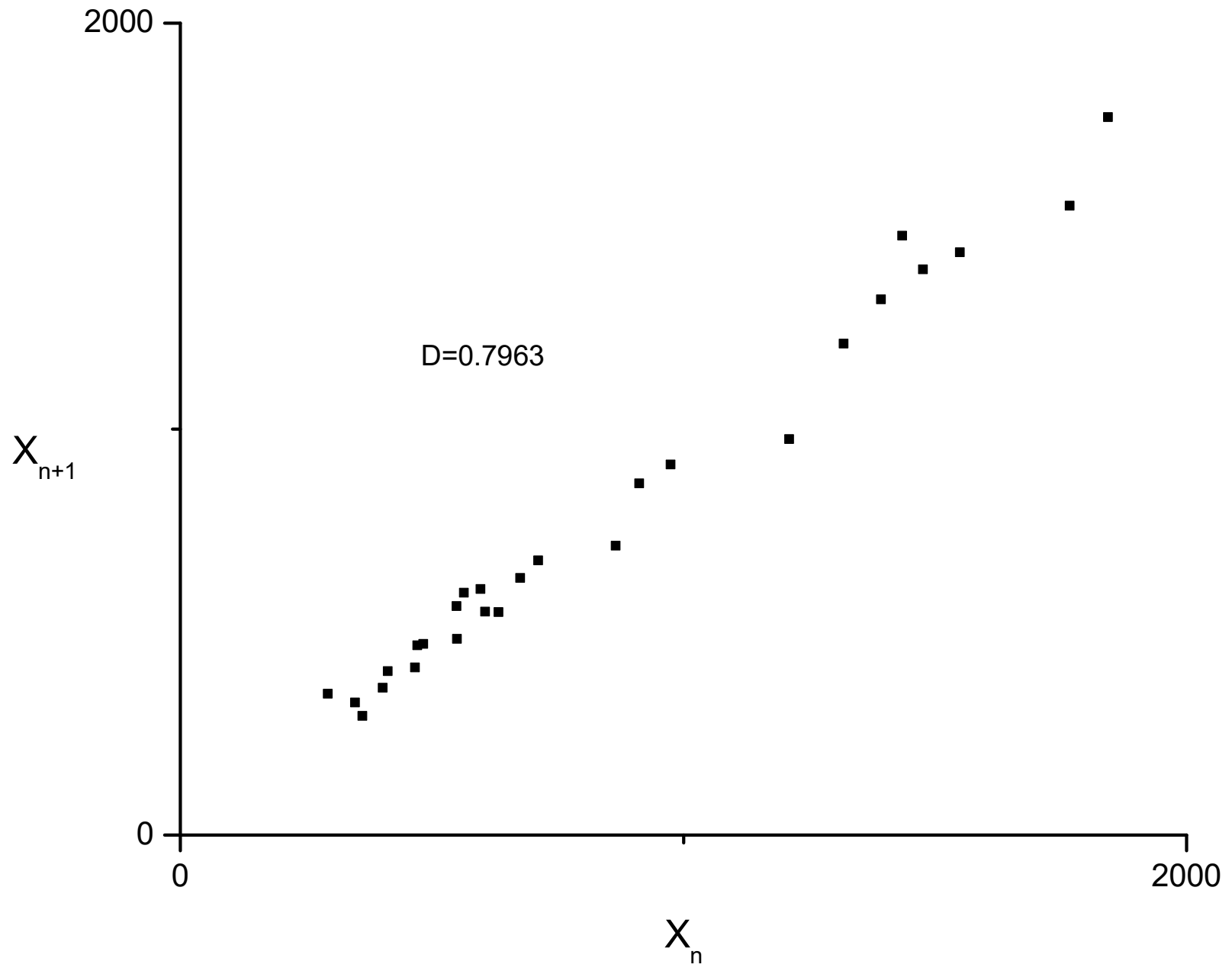

Fig. 11.- Sección de Poincaré para los últimos casos, desde el 2/8/21 al 15/11/21. La dimensión fractal del mapa en este caso es 0.7963.

Se ve que las características caóticas se han debilitado. Podemos comparar con el comportamiento de los primeros 200 días, en que la dimensión fractal era mayor que 1, y se observa una disminución notable

## VIII.- CONCLUSIONES

Se comprueba aquí la naturaleza holística y multifactorial del sistema creado en Cuba por la propagación del COVID-19. El comportamiento crítico auto organizado (es decir, cuando el sistema tiende a un estado con correlaciones de largo alcance y dinámica caótica débil) se ha visto en fechas reciente influido por factores que debilitan las correlaciones y tienden a actuar como un "sintonizador" que dificulta la dinámica anterior. Es de esperar entonces que la dimensión fractal del mapa de Poincaré cambie y no se alcance el valor asintótico predicho por el comportamiento inicial. La alometría entre provincias es de esperar que se rompa y se llegue a un estado con correlaciones débiles. A nuestro entender, lo único que ha cambiado esencialmente es la presencia de la vacunación, que ha provocado un cambio cualitativo en los componentes del sistema.

No obstante, estas conclusiones no son válidas si no se tiene en cuenta la próxima influencia que puede debilitar este proceso: La influencia de nuevas acciones de tipo administrativo, de protocolo, biológico, que puedan eliminar el apantallamiento introducido por las vacunas.

Como ejemplo podemos notar que, si bien la presencia de alometría entre provincias en las distintas fechas demuestra la presencia de correlaciones de largo alcance y la presencia de complejidad en Cuba, lo mismo puede hacerse para el resto del mundo como un todo y juzgar sobre su estado actual.

La figura 14 muestra los datos de número de casos presentados acorde con la información diaria del sitio Worldometer (https://www.worldometer.info). Para el día 28/10/2021:

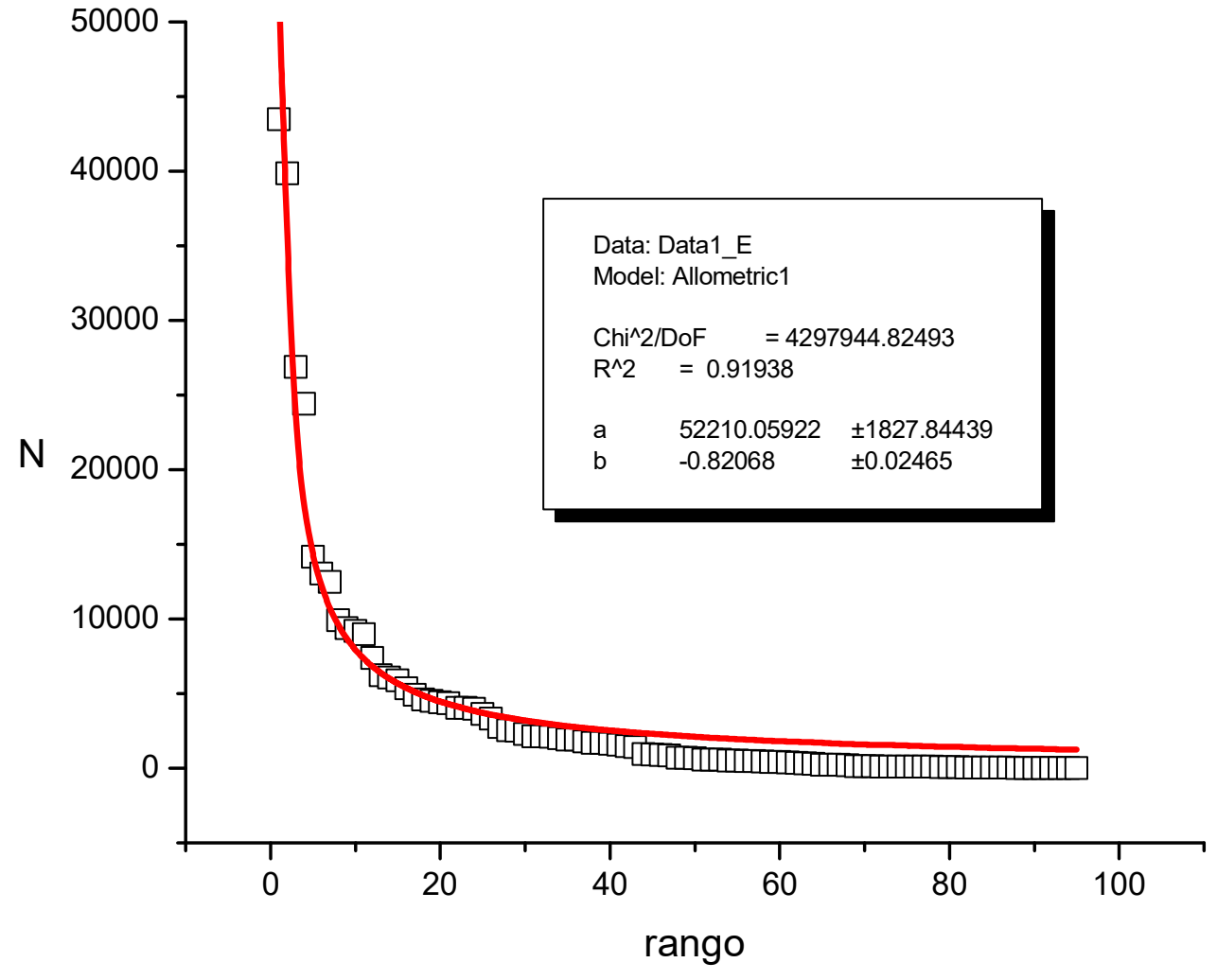

Fig.12.-Alometría de los casos diarios el día 28/10/2021. La ley de escala es evidente (ver R^2) La data pertenece a los 96 países que informaron ese día.

Más convincente es aún la distribución de los fallecimientos en el mundo. Para ese mismo día los fallecimientos de 90 países que informaron cumplen también una ley de escala con un coeficiente de correlación que muestra una coincidencia extraordinaria entre la data y la ley de potencia.:

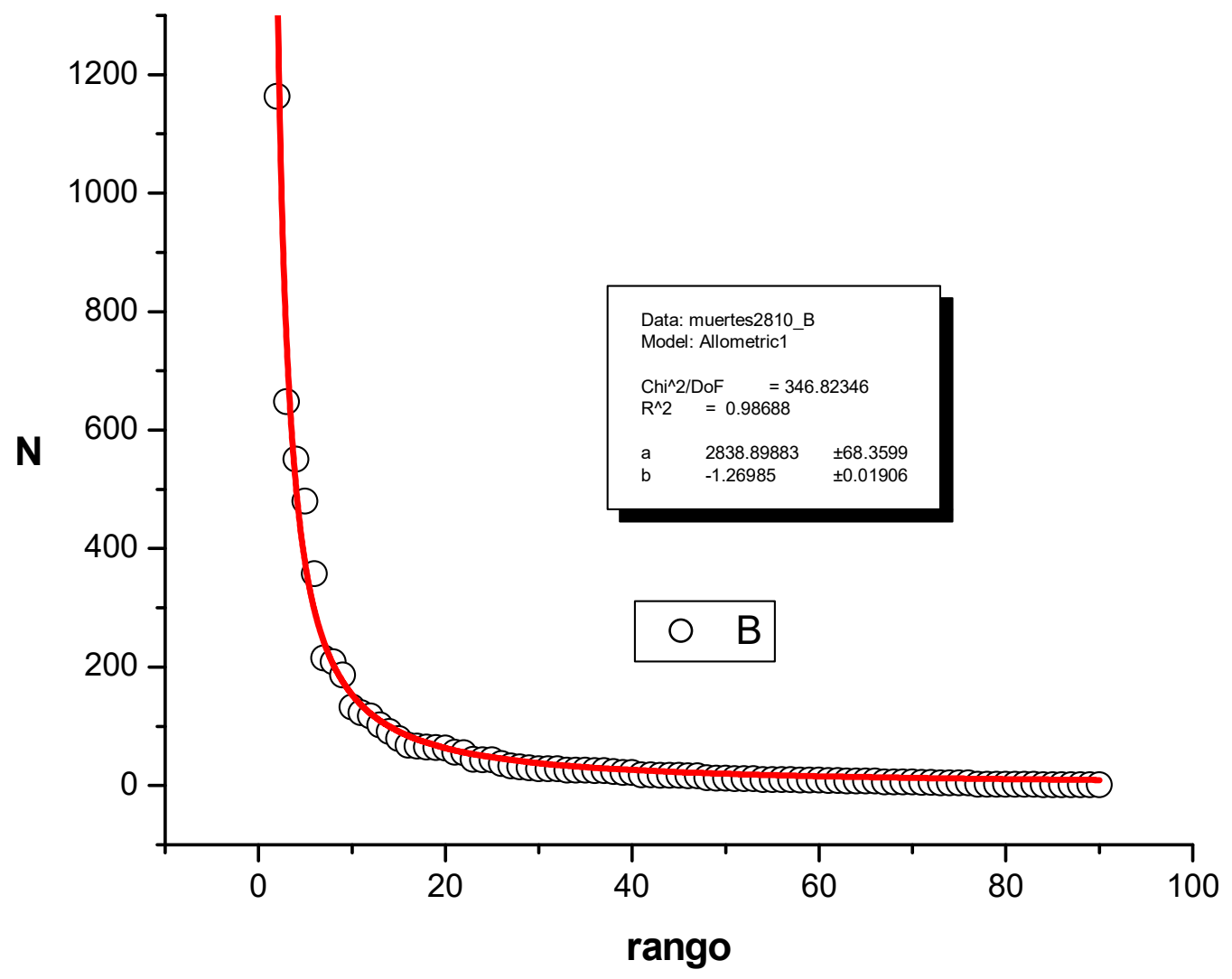

Fig.12.-Alometría de los fallecimientos diarios el día 28/10/2021. La ley de escala es evidente (ver R^2) La data pertenece a los 90 países que informaron ese día. Aquí vale la pena citar: "La muerte nos lleva el dedo por sobre el libro de la vida" (José Martí, "El presidio político en Cuba", Madrid, 1871)

De lo mostrado parece obvio que paralelamente a la eliminación de las medidas de vacunación, aislamiento y limitación de movilidad que se vayan instrumentando, se dirijan acciones a la eliminación de la posibilidad de surgimiento de algún nuevo factor generador de caos. Esto debe ser objeto de análisis multifactorial.

## BIBLIOGRAFIA